\documentclass[a4paper,10pt,onecolumn,oneside,notitlepage,final]{article}
\usepackage{graphics}

 \mathchardef\ScriptA="7241
 \mathchardef\ScriptB="7242 
 \mathchardef\ScriptC="7243
 \mathchardef\ScriptD="7244
 \mathchardef\ScriptE="7245
 \mathchardef\ScriptF="7246
 \mathchardef\ScriptG="7247
 \mathchardef\ScriptH="7248
 \mathchardef\ScriptI="7249
 \mathchardef\ScriptJ="724A
 \mathchardef\ScriptK="724B
 \mathchardef\ScriptL="724C
 \mathchardef\ScriptM="724D
 \mathchardef\ScriptN="724E
 \mathchardef\ScriptO="724F
 \mathchardef\ScriptP="7250
 \mathchardef\ScriptQ="7251
 \mathchardef\ScriptR="7252
 \mathchardef\ScriptS="7253
 \mathchardef\ScriptT="7254
 \mathchardef\ScriptU="7255
 \mathchardef\ScriptV="7256
 \mathchardef\ScriptW="7257
 \mathchardef\ScriptX="7258
 \mathchardef\ScriptY="7259
 \mathchardef\ScriptZ="725A

\def\mapgeq{\mathbin{\lower.3ex\hbox{$\buildrel>\over{\smash{\scriptstyle\sim}\vphantom{_x}}$}}}
\def\mapleq{\mathbin{\lower.3ex\hbox{$\buildrel<\over{\smash{\scriptstyle\sim}\vphantom{_x}}$}}}
\def\mapgeqeq{\mathbin{\lower.3ex\hbox{$\buildrel>\over{\smash{\scriptstyle\approx}\vphantom{_2}}$}}}
\def\mapleqeq{\mathbin{\lower.3ex\hbox{$\buildrel<\over{\smash{\scriptstyle\approx}\vphantom{_2}}$}}}

 \mathchardef\#="0023
 \mathchardef\$="0024
 \mathchardef\%="0025
 \mathchardef\ddash="705C
 
 \mathchardef\lwavy="336E
 \mathchardef\rwavy="336F
 \mathchardef\biglwavy="331A
 \mathchardef\bigrwavy="331B
 \mathchardef\bigglwavy="3328
 \mathchardef\biggrwavy="3329
 \mathchardef\littlesum="0350

\begin{document}
\title{
\Large \bf Neutrino Oscillations Induced by Two-loop Radiative Effects
}
\author{{\normalsize Teruyuki Kitabayashi$^a$} 
            \footnote{E-mail:teruyuki@post.kek.jp}
            \hspace{2mm} and
        {\normalsize Masaki Yasu\`{e}$^b$}
            \footnote{E-mail:yasue@keyaki.cc.u-tokai.ac.jp}
            \footnote{A talk given at {\it Post Summer Institute 2000 
            on Neutrino Physics}, August 21-24, 2000, Fuji-Yoshida, Japan.}
\vspace{2mm}
        \\
        $^a$
		{\small \sl Accelerator Engineering Center,}\\ 
        {\small \sl Mitsubishi Electric System \& Service Engineering Co.Ltd.,} \\
        {\small \sl 2-8-8 Umezono, Tsukuba, Ibaraki 305-0045, Japan.}\\
        \\
        $^b$
        {\small \sl Department of Natural Science,}\\
        {\small \sl School of Marine Science and Technology, Tokai University,}\\
        {\small \sl 3-20-1 Orido, Shimizu, Shizuoka 424-8610, Japan}
        \\
        {\small \sl and}
        \\
        {\small \sl Department of Physics, Tokai University,}\\
        {\small \sl 1117 Kitakaname, Hiratsuka, Kanagawa 259-1291, Japan.}
        }

\date{\small TOKAI-HEP/TH-0008, October, 2000}
\maketitle

\begin{abstract}
Phenomena of neutrino oscillations are discussed on the basis of two-loop radiative neutrino mechanism.  Neutrino mixings are experimentally suggested to be maximal in both atmospheric and solar neutrino oscillations.  By using $L_e - L_\mu - L_\tau$ ($\equiv L^\prime$)-conservation, which, however, only ensures the maximal solar neutrino mixing, we find that two-loop radiative mechanism dynamically generates the maximal atmospheric neutrino mixing and that the estimate of $\Delta m^2_\odot/\Delta m^2_{atm} \sim\epsilon  m_e/m_\tau$ explains $\Delta m^2_\odot/\Delta m^2_{atm} \ll 1$ because of $m_e/m_\tau \ll 1$, where $\epsilon$ measures the breaking of the $L^\prime$-conservation.  Together with $\Delta m^2_{atm} \approx 3\times 10^{-3}$ eV$^2$, this estimate yields $\Delta m^2_\odot \sim 10^{-7}$ eV$^2$ for $\epsilon \sim 0.1$, which corresponds to the LOW solution to the solar neutrino problem.  Neutrino mass scale is given by $(16\pi^2)^{-2} m_em_\tau /M$ ($M \sim$ 1 TeV), which is of order 0.01 eV.
\end{abstract}

\vspace{2mm}
\section{Introduction}
Neutrino oscillations have been long recognized if neutrinos are massive particles \cite{EarlyMassive}.  Such oscillations in fact have been recently confirmed by the Super-Kamiokande collaboration \cite{SuperKamiokande} and have also been observed for solar neutrinos produced inside the Sun \cite{SolarNeutrino}.  The recent report from the K2K collaboration \cite{RecentK2K} has further shown that the atmospheric neutrino oscillations are characterized by $\Delta m^2_{atm} \approx 3 \times 10^{-3}$ eV$^2$, which implies $\sim$5.5$\times 10^{-2}$ eV as neutrino masses.  This tiny mass scale for neutrinos can be generated by radiative mechanisms, where the smallness originates from the smallness of radiative effects \cite{Zee,Babu}.  Radiative mechanisms uses $L$=2 interactions given by $\nu^{[i}_L\ell^{j]}_L$ for one-loop radiative effects \cite{Zee,Radiative,ZeeType,Usefulness} and by additional $\ell^{\{i}_R\ell^{j\}}_R$ for two-loop radiative effects \cite{Babu,2LoopOnly}, where $i$ and $j$ denote three families ($i$,$j$ = 1,2,3). 

\begin{figure}[t]
  \begin{center}
    \includegraphics*[30mm,230mm][190mm,280mm]{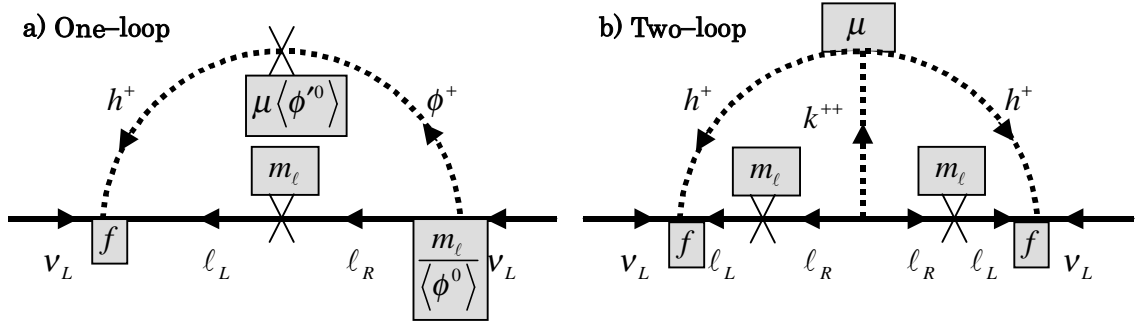}
    \caption{\label{fig1}
	Radiatively generated Majorana neutrino masses: 
    (a) one-loop diagram,
    (b) two-loop diagram.}
  \end{center}
\end{figure}

At the one-loop level, Zee \cite{Zee} has presented the mechanism that utilizes a new standard Higgs scalar called $\phi^\prime$ in addition to the standard Higgs scalar, $\phi$, both of which are $SU(2)_L$-doublets, and another singly charged scalar called $h^+$, which is an  $SU(2)_L$-singlet, with the coupling of $f_{[ij]}\nu^{[i}_L\ell^{j]}_Lh^+$. The Fermi statistics forces $\nu^{[i}_L\ell^{j]}_L$ to be antisymmetrized with respect to the family indices. After the spontaneous breakdown of $SU(2)_L \times U(1)_Y$, an interaction of $\phi\phi^\prime h^+$ yields the possible mixing of $h^+$ with $\phi^+$ characterized by the scale of $\mu$, which finally induces Majorana neutrino masses.  Again, the Fermi statistics forces $\phi\phi^\prime$ to be antisymmetrized with respect to the $SU(2)_L$-indices. Depicted in Figure \ref{fig1}(a) is the diagram for generating Majorana neutrino masses. 
The order-of-magnitude estimate gives the one-loop neutrino mass, $m^{1-loop}_\nu$, for $\nu_i$-$\nu_j$ to be:
\begin{eqnarray}
&& m^{1-loop}_\nu \sim f_{[ij]}\frac{m^2_{\ell^j}}{16\pi^2M^2}\mu,
\label{Eq:OneLoopNuMass}
\end{eqnarray}
for $\langle 0 | \phi^0 | 0 \rangle$ $\sim$ $\langle 0 | \phi^{\prime 0} | 0 \rangle$, where $M$ stands for the scale of the model, presumably of order 1 TeV.  The factor of $16\pi^2$ in the denominator is specific to one-loop radiative corrections.  This estimate turns out to be 
\begin{eqnarray}
&& m^{1-loop}_\nu \sim 2 \times 10^3 f_{[i\tau ]}~(\frac{\mu}{100~{\rm GeV}})~{\rm eV},
\label{Eq:OneLoopNuMassNumerics}
\end{eqnarray}
for $m_{\ell^j}$ = $m_\tau$ ($j$=$\tau$).  To obtain $m_\nu$ $\sim$ 0.1 eV, we require that
\begin{eqnarray}
&& f_{[i\tau ]} \sim 5 \times 10^{-5},
\label{Eq:OneLoopNuMassCoupling}
\end{eqnarray}
for $\mu$ $\sim$ 100 GeV. 
Therefore, to get tiny neutrino masses of order 0.1 eV, one has to give excessive suppression to the lepton-number violating $\nu\ell$-coupling.

At the two-loop level, additional suppression arises.  In addition to $h^+$, a doubly charged $k^{++}$-scalar is required to realize the mechanism of the Zee-Babu type \cite{Babu} and $k^{++}$ couples to a right-handed charged lepton pair via $\ell^{\{i}_R\ell^{j\}}_Rk^{++}$ with coupling strength of $f_{\{ij\}}$.  Using a possible coupling of this new $k^{++}$ with $h^+$ via $h^+h^+k^{++\dagger}$, we can find interactions corresponding to Figure \ref{fig1}(b). The order-of-magnitude estimate gives the two-loop neutrino mass, $m^{2-loop}_\nu$, for $\nu_i$-$\nu_k$ to be:
\begin{eqnarray}
&& m^{2-loop}_\nu \sim f_{[ij]}f_{\{jj^\prime\}}f_{[kj^\prime]}\frac{m_{\ell^j}m_{\ell^{j^\prime}}}{(16\pi^2)^2M^2}\mu.\label{Eq:TwoLoopNuMass}
\end{eqnarray}
The factor of $(16\pi^2)^2$ in the denominator is specific to two-loop radiative corrections.  This estimate turns out to yield 
\begin{eqnarray}
&& m^{2-loop}_\nu \sim 10 f_{[i\tau ]}f_{[\tau\tau ]}f_{[j\tau ]}~(\frac{\mu}{100~{\rm GeV}})~{\rm eV},
\label{Eq:TwoLoopNuMassNumerics}
\end{eqnarray}
for $m_{\ell^j,\ell^{j^\prime}}$ = $m_\tau$ ($j,j^\prime$=$\tau $).  To obtain $m_\nu$ $\sim$ 0.1 eV, thanks to the extra loop-factor of $16\pi^2$, we only require that
\begin{eqnarray}
&& f_{[i\tau ]} \sim 0.1,
\label{Eq:TwoLoopNuMassCoupling}
\end{eqnarray}
for $f_{[\tau\tau ]} \sim 1$ and $\mu$ $\sim$ 100 GeV. Therefore, the two-loop radiative neutrino masses can be of order of 0.1 eV without excessive suppression for relevant couplings \cite{TwoLoopGood}.

\section{Bimaximal Mixing}
The observed pattern of neutrino oscillations is consistent with the pattern arising from the requirement of the conservation of the new quantum number $L_e-L_\mu-L_\tau$ ($\equiv L^\prime$) \cite{LeLmuLtau}.
The $U(1)_{L^\prime}$ symmetry based on the $L^\prime$-conservation can be used to describe the bimaximal mixing scheme for neutrino oscillations \cite{BiMaximal,NearlyBiMaximal}.  However, the $L^\prime$-conservation itself only ensures the maximal solar neutrino mixing but does not determine the atmospheric neutrino mixing angle.  In fact, in the one-loop radiative mechanism, fine-tuning of lepton-number violating couplings is necessary to yield bimaximal mixing for atmospheric neutrino oscillations. 

In the one-loop radiative mechanism, we have known the form of the neutrino mass matrix, which is given by
\begin{eqnarray}
&& M_\nu   \propto \left. {\left( {\begin{array}{*{20}c}
   0 & {f_{[e\mu ]} m_\mu ^2 } & {f_{[e\tau ]} m_\tau ^2 }  \\
   {} & 0 & {f_{[\mu \tau ]} m_\tau ^2 }  \\
   {} & {} & 0  \\
\end{array}} \right)} \right|_{sym}  \Rightarrow \left( {\begin{array}{*{20}c}
   0 & { \sim 1} & { \sim 1}  \\
   {} & 0 & {\varepsilon \left( { \ll 1} \right)}  \\
   {} & {} & 0  \\
\end{array}} \right)m ,
\label{Eq:OneLoopNuMassMatrix}
\end{eqnarray}
where $m$ stands for the neutrino mass scale.  The bimaximal mixing is realized if the couplings satisfy 
\begin{eqnarray}
&& f_{e\mu } m_\mu ^2  = f_{[e\tau ]} m_\tau ^2  \Rightarrow f_{[e\mu ]}  \gg f_{[e\tau ]} \left( { \gg f_{[\mu \tau ]}  \approx 0} \right),
\label{Eq:OneLoopNuCouplings}
\end{eqnarray}
indicating the fine-tuning of the couplings $f$'s.  This fine-tuning is referred to as $\ddash$inverse hierarchy in the couplings", namely, $f_{[e\mu]} \gg f_{[e\tau]}$ \cite{InverseHierarchy}. 
The $L^\prime$-conservation gives $f_{[\mu\tau]}$=0.  Its tiny breaking effect characterized by the parameter, $\varepsilon$, produces tiny solar neutrino oscillations.  

On the other hand, in the two-loop radiative mechanism, we will find the mass matrix \cite{2LoopOnly} given by
\begin{eqnarray}
&& M_\nu   \propto \left. {\left( {\begin{array}{*{20}c}
   0 & {f_{[e\tau ]} f_{[e\mu ]} m_e m_\tau  } & {f_{[e\tau ]} f_{[e\tau ]} m_e m_\tau  }  \\
   {} & {f_{[e\mu ]} f_{[e\mu ]} m_e^2 } & {f_{[e\mu ]} f_{[e\tau ]} m_e^2 }  \\
   {} & {} & {f_{[e\tau ]} f_{[e\tau ]} m_e^2 }  \\
\end{array}} \right)} \right|_{sym}  \Rightarrow \left( {\begin{array}{*{20}c}
   0 & { \sim 1} & { \sim 1}  \\
   {} & \varepsilon  & {\varepsilon '}  \\
   {} & {} & {\varepsilon ''}
\end{array}} \right)m.
\label{Eq:TwoLoopNuMassMatrix}
\end{eqnarray}
The bimaximal structure is reproduced if
\begin{eqnarray}
&& f_{[e\tau ]} f_{[e\mu ]} m_e m_\tau   = f_{[e\tau ]} f_{[e\tau ]} m_e m_\tau   \Rightarrow f_{[e\mu ]}=f_{[e\tau ]}.
\label{Eq:TwoLoopNuCouplings}
\end{eqnarray}
Therefore, no hierarchy in the couplings is necessary. The breaking of the $L^\prime$-conservation gives the suppressed entries, $\varepsilon$, $\varepsilon^\prime$, $\varepsilon^{\prime\prime}$,  proportional to $m_e^2 $. Therefore, we observe that
\begin{eqnarray}
&& {{\Delta m_ \odot ^2 } \mathord{\left/
 {\vphantom {{\Delta m_ \odot ^2 } {\Delta m_{atm}^2 }}} \right.
 \kern-\nulldelimiterspace} {\Delta m_{atm}^2 }} \propto {{m_e } \mathord{\left/
 {\vphantom {{m_e } {m_\tau  }}} \right.
 \kern-\nulldelimiterspace} {m_\tau  }},
\label{Eq:TwoLoopCouplings}
\end{eqnarray}
which dynamically guarantees $\Delta m_{atm}^2 \gg \Delta m_ \odot ^2$ because of $m_\tau \gg m_e$.  

In radiative mechanisms, the hierarchy of $\Delta m_{atm}^2 \gg \Delta m_ \odot ^2$ can also be ascribed to the generic smallness of two-loop radiative effects over one-loop radiative effects \cite{2vs1Loop}. Therefore, we have in hands two dynamical reasons for $\Delta m_{atm}^2 \gg \Delta m_ \odot ^2$:  
\begin{eqnarray}
&& \frac{{\Delta m_ \odot ^2 }}{{\Delta m_{atm}^2 }} \ll 1~{\rm  because }\left\{ \begin{array}{l}
 {{{\rm 2 - loop}} \mathord{\left/
 {\vphantom {{{\rm 2 - loop}} {{\rm 1 - loop}}}} \right.
 \kern-\nulldelimiterspace} {{\rm 1 - loop}}} \ll 1\\ 
 {{m_e } \mathord{\left/
 {\vphantom {{m_e } {m_\tau  }}} \right.
 \kern-\nulldelimiterspace} {m_\tau  }} \ll 1 \\ 
 \end{array} \right. .
\label{Eq:MassHierarchy}
\end{eqnarray}

\section{Two-loop Radiative Neutrino Masses}
Interactions that we introduce can be classified by the ordinary lepton number ($L$) and $L^\prime$-number of particles, which are listed in the Table 1.  
\begin{table}[h]
\caption{
    \label{TabLnumberNM}
    $L$ and $L^\prime$ quantum numbers.}
\begin{center}
\begin{tabular}{|c|cccccc|}
    \hline
        Fields & $(\nu_{eL}, e^-_L), e^-_R$ & $(\nu_{iL}, \ell^-_{iL}), \ell^-_{iR}\vert_{i=\mu,\tau}$ & $\phi$
               & $h^+$ & $k^{++}$  & $k^{\prime ++}$\\
    \hline
        $L$        & 1 & 1 &  0 & $-$2  & $-$2 & $-$2\\    
    \hline
        $L^\prime$ & 1 & $-$1 &  0 & 0  & 0 & $-$2\\
    \hline
\end{tabular}
\end{center}
\end{table}
The new ingredients that are not contained in the standard model are the $SU(2)_L$-singlet scalars, $h^+$ and $k^{++}$.  We have further employed an additional $k^{++}$ to be denoted by $k^{\prime++}$ in order to import the $L^\prime$- breaking. The $L$- and $L^\prime$-quantum number of $k^{\prime ++}$ is also listed in Table 1. Extra $L$- and $L^\prime$-conserving Yukawa interactions are given by
\begin{eqnarray}
& \left\{ \begin{array}{l}
 f_{[ej]} \left( {\nu _{eL} \ell ^j_L   - \nu^j _L e_L^ -  } \right)h^ + ,  \\ 
 f_{\{ ej\} } e_R^ -  \ell^j _R  k^{ +  + },  \\ 
 \frac{1}{2}f_{\{ ee\} } e_R^ -  e_R^ -  k'^{ +  + }.  \\ 
 \end{array} \right.
\label{Eq:TwoLoopInt}
\end{eqnarray}
An $L$-breaking but $L^\prime$-conserving interaction is specified by
\begin{eqnarray}
&& \mu _0 h^ +  h^ +  k^{ +  + \dagger },
\label{Eq:TwoLoopInt2}
\end{eqnarray}
where $\mu_0$ represents a mass scale. An $L^\prime$-breaking interaction is activated by $k^{\prime ++}$ via
\begin{eqnarray}
&& \mu _b h^ +  h^ +  k'^{ +  + \dagger },
\label{Eq:TwoLoopInt3}
\end{eqnarray}
where $\mu_b$ represents a breaking scale of the $L^\prime$-conservation. 

Yukawa interactions, then, take the form of 
\footnote{
The corresponding expression of Eq.(2) in Ref.\cite{Details} should read this equation.
}
\begin{eqnarray}
-{\cal L}_Y  & = &  
\sum_{i=e,\mu,\tau}
f^i_\phi {\overline {\psi^i_L}}\phi\ell^i_R
+
\sum_{i=\mu,\tau}
\left(
f_{[ei]}{\overline {\left( \psi_L^e \right)^c}}\psi^i_Lh^+
+f_{\{ei\}}{\overline {\left( e_R \right)^c}}\ell^i_Rk^{++} 
\right)
\nonumber \\
& &+ 
 \frac{1}{2}f_{\{ee\}}{\overline {\left( e_R \right)^c}}e_Rk^{\prime ++} 
+ {\rm (h.c.)},
\label{Eq:OurYukawa}
\end{eqnarray}
and Higgs interactions are described by
self-Hermitian terms composed of $\varphi\varphi^\dagger$ ($\varphi$ = $\phi$, $h^+$, 
$k^{++}$, $k^{\prime ++}$) and by the non-self-Hermitian terms in
\begin{eqnarray}\label{Eq:Conserved}
V_0 &=&  
\mu_0h^+h^+k^{++^\dagger}
+ {\rm (h.c.)}.
\end{eqnarray}
This coupling softly breaks the $L$-conservation 
but preserves the $L^\prime$-conservation.  
To account for solar neutrino oscillations, the breaking of the $L^\prime$-conservation 
should be included and is assumed to be furnished by 
\begin{eqnarray}\label{Eq:Broken}
V_b & = &\mu_bh^+h^+k^{\prime ++\dagger} + {\rm (h.c.)}.
\end{eqnarray}
 
\begin{figure}[t]
  \begin{center}
    \includegraphics*[30mm,229mm][190mm,280mm]{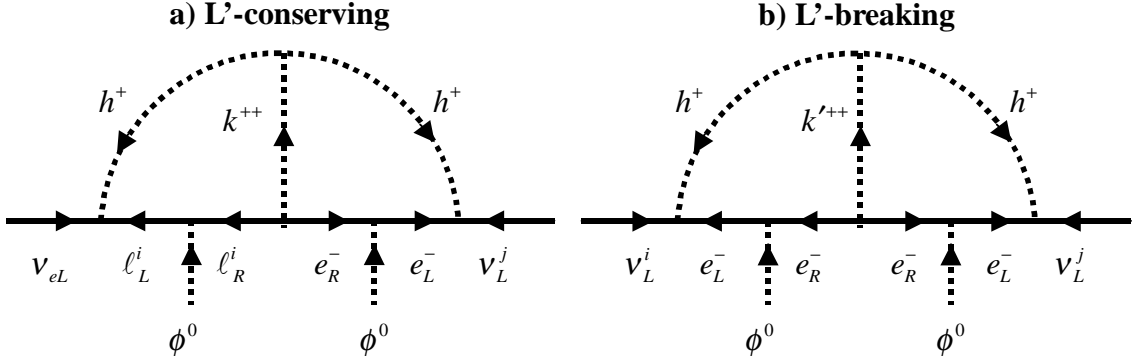}
    \caption{\label{fig2}
	Radiatively generated Majorana neutrino masses: 
    (a) $L^\prime$-conserving two-loop diagram,
    (b) $L^\prime$-breaking two-loop diagram.}
  \end{center}
\end{figure}

Neutrino masses are generated by interactions corresponding to the diagrams depicted in Figure \ref{fig2}(a,b).
The resulting Majorana neutrino mass matrix is given by
\begin{eqnarray}
&& M_\nu   = \left( {\begin{array}{*{20}c}
   0 & {m_{e\mu } } & {m_{e\tau } }  \\
   {m_{e\mu } } & {\delta _{\mu \mu } } & {\delta _{\mu \tau } }  \\
   {m_{e\tau } } & {\delta _{\mu \tau } } & {\delta _{\tau \tau } }  \\
\end{array}} \right).
\label{Eq:TinyMassEntries}
\end{eqnarray}
Here, the bimaximal structure is controlled by
\begin{eqnarray}
&& m_{ei}  \approx -2f_{[e\tau ]} f_{[ei]} f_{\{ \tau e\} } \frac{{m_\tau  m_e }}{{m_k^2 }}\mu _0 \left[ {\frac{1}{{16\pi ^2 }}\ln \left( {\frac{{m_k^2 }}{{m_h^2 }}} \right)} \right]^2~(i = \mu, \tau ),
\label{Eq:TinyMassEntries_ei}
\end{eqnarray}
where the product of $m_e$ and $m_\tau$ appears.  This is because the exchanged leptons are $e$ and $\tau$ as can been seen from Figure \ref{fig2}(a).  Tiny splitting is induced by
\begin{eqnarray}
&& \delta _{ij}  \approx  - f_{[ei]} f_{[ej]} f_{\{ ee\} } \frac{{m_e m_e }}{{m_{k'}^2 }}\mu _b \left[ {\frac{1}{{16\pi ^2 }}\ln \left( {\frac{{m_{k'}^2 }}{{m_h^2 }}} \right)} \right]^2,
\label{Eq:TinyMassEntries_ij}
\end{eqnarray}
where $m^2_e$ appears because the exchanged leptons are both $e$ and $e$ as can been seen from Figure \ref{fig2}(b). These expressions, Eqs.(\ref{Eq:TinyMassEntries_ei}) and (\ref{Eq:TinyMassEntries_ij}), are subject to the approximation of  $m^2_{k,k^\prime}\gg$(other mass squared).  The detailed derivation of Eqs.(\ref{Eq:TinyMassEntries_ei}) and (\ref{Eq:TinyMassEntries_ij}) can be found in the Appendix of Ref.\cite{Details}.
Oscillations are described by these mass parameters:
\begin{eqnarray}
&& \Delta m_{atm}^2  = m_{e\mu }^2  + m_{e\tau }^2 \left( { \equiv m_\nu ^2 } \right), \quad \Delta m_ \odot ^2  = 4m_\nu  \delta m,
\label{Eq:OscMasses}
\end{eqnarray}
where
\begin{eqnarray}
&& \delta m = \frac{1}{2}\left| {\delta _{\mu \mu } \cos ^2 \theta _\nu   + 2\delta _{\mu \tau } \cos \theta _\nu  \sin \theta _\nu   + \delta _{\tau \tau } \sin ^2 \theta _\nu  } \right| 
\label{Eq:OscParameters}
\end{eqnarray}
with
\begin{eqnarray}
&& \cos \theta _\nu   = m_{e\mu } /m_\nu  , \quad
\sin \theta _\nu   = m_{e\tau } /m_\nu. 
\label{Eq:OscAngle}
\end{eqnarray}

It is thus found that (nearly) bimaximal mixing is reproduced by requiring
\begin{eqnarray}
&& f_{[e\mu ]}  \approx f_{[e\tau ]},
\label{Eq:NaturalCoupling}
\end{eqnarray}
yielding $\sin 2\theta_\nu \approx 1$. Tiny mass-splitting $\Delta m_{atm}^2 \gg \Delta m_ \odot ^2 $ is ensured by the mass-hierarchy:
\begin{eqnarray}
&& m_\tau  \gg m_e.
\label{Eq:LeptoHierarchy}
\end{eqnarray}
As a result, we obtain an estimate of the ratio:
\begin{eqnarray}
&& \frac{{\Delta m_ \odot ^2 }}{{\Delta m_{atm}^2 }} \sim \frac{{\mu _b }}{{\mu _0 }}\frac{{m_e }}{{m_\tau  }}\frac{{m_k^2 }}{{m_{k'}^2 }}.
\label{Eq:MainResult}
\end{eqnarray}
From this estimate, we find that 
\begin{eqnarray}
&& \Delta m_ \odot ^2  \sim 3 \times 10^{ - 4} \frac{{\mu _b }}{{\mu _0 }}\Delta m_{atm}^2 ~\left( {m_k^2  \sim m_{k'}^2 } \right)
\nonumber \\
&& \Rightarrow \Delta m_ \odot ^2  \sim 3 \times 10^{ - 5} \Delta m_{atm}^2 ~\left( {\mu _b  \sim {{\mu _0 } \mathord{\left/{\vphantom {{\mu _0 } {10}}} \right.
 \kern-\nulldelimiterspace} {10}}} \right)
\nonumber \\
&& \Rightarrow \Delta m_ \odot ^2  \sim 10^{ - 7}~{\rm  eV}^{\rm 2} ~\left( {\Delta m_{atm}^2  \sim 3 \times 10^{ - 3} {\rm  eV}^{\rm 2} } \right).
\label{Eq:Numerics}
\end{eqnarray}
The resulting $\Delta m_ \odot ^2$ corresponds to the allowed region for the LOW solution to the solar neutrino problem.
Since $k^{++}$ and $k^{\prime ++}$ couple to the charged lepton pairs, these scalars produce extra contributions on the well-established low-energy phenomenology.  In particular, we should consider effects from $\mu^- \to e^- \gamma$, $e^-e^-e^+$, $e^-e^- \to e^-e^-$ and $\nu_\mu e^- \to \nu_\mu e^-$.  
\begin{figure}[t]
  \begin{center}
    \includegraphics*[30mm,230mm][190mm,280mm]{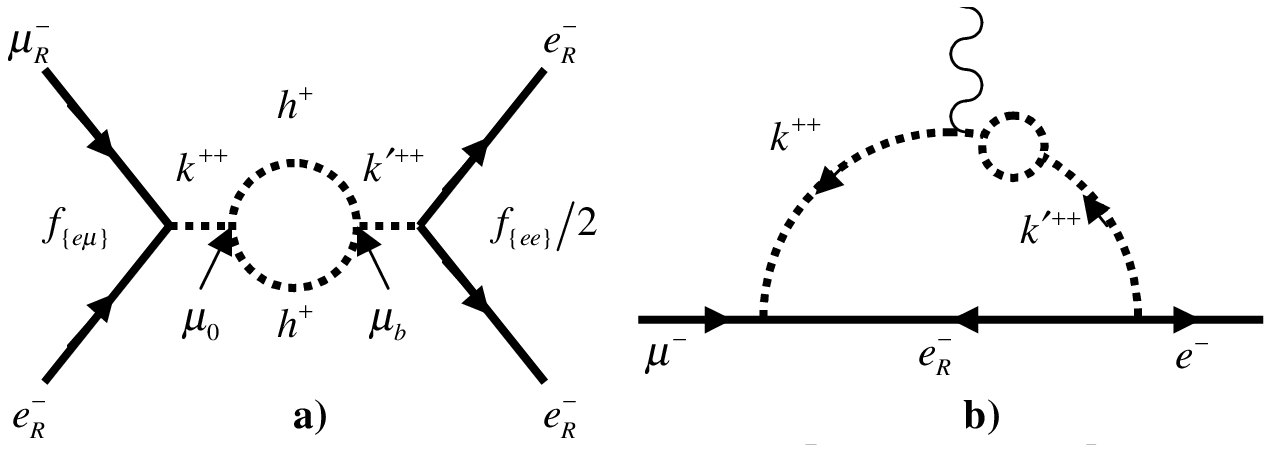}
    \caption{\label{fig3}
    (a) $\mu^- \to e^-e^-e^+$,
    (b) $\mu^- \to e^- \gamma$.}
  \end{center}
\end{figure}
\begin{figure}[t]
  \begin{center}
    \includegraphics*[30mm,230mm][190mm,280mm]{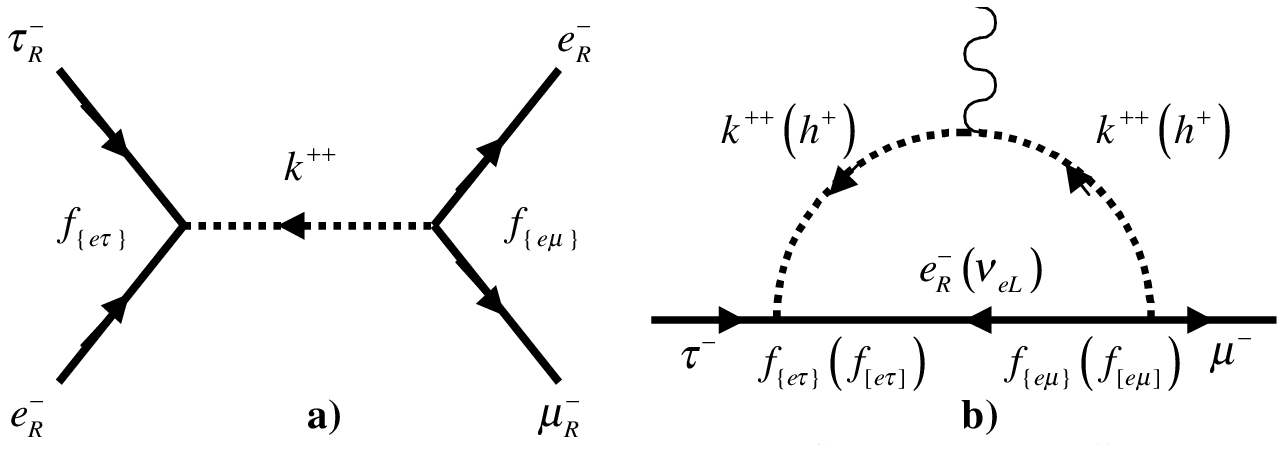}
    \caption{\label{fig4}
    (a) $\tau^- \to \mu^-e^-e^+$,
    (b) $\tau^- \to \mu^- \gamma$.}
  \end{center}
\end{figure}
\begin{figure}[t]
  \begin{center}
    \includegraphics*[30mm,230mm][190mm,280mm]{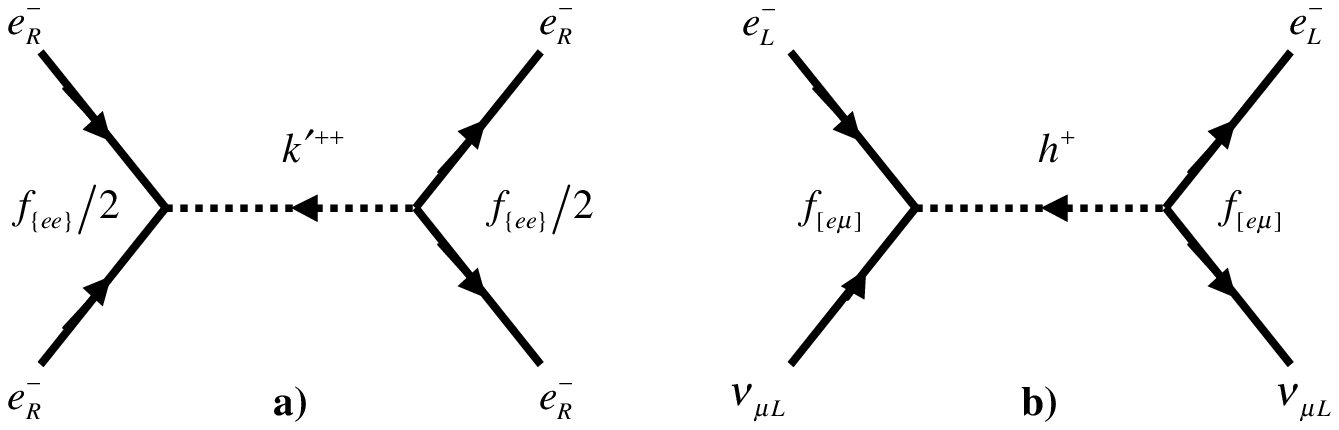}
    \caption{\label{fig5}
    (a) $e^-e^- \to e^-e^-$,
    (b) $\nu_\mu e^- \to \nu_\mu e^-$.}
  \end{center}
\end{figure}
The relevant constraints on the parameters associated with the scalars of $h^+$, $k^{++}$ and $k^{\prime ++}$ are, thus, given by
\footnote{
The constraints of Eqs.(12) and (13) in Ref.\cite{Details} should, respectively, be replaced by the corresponding bounds in the items 1, 3 and 4.  Namely, $f_{\{11,12\}}$ should read $f_{\{11,12\}}/2$ in Ref.\cite{Details}.
}
\begin{enumerate}
\item $\mu ^ -   ~\to~e^ -  e^ -  e^ + $ in Figure \ref{fig3}(a) and $\mu ^ -  ~\to~e^ -   \gamma$ in Figure \ref{fig3}(b) \cite{Mu_E} (forbidden by the $L^\prime$-conservation), yielding
\begin{eqnarray}
&& \frac{{\xi f_{\{ e\mu \} } f_{\{ ee\} } }}{{\bar m_k^2 }} < \left\{ \begin{array}{l}
 1.2 \times 10^{ - 10}~{\rm  GeV}^{ - 2}~{\rm from}~B\left( {\mu ^ -   \to e^ -  e^ -  e^ +  } \right) < 10^{-12}~\cite{Data}\\ 
 2.4 \times 10^{ - 8}~{\rm  GeV}^{ - 2}~{\rm from}~B\left( {\mu ^ -   \to e^ -   \gamma } \right) < 1.2 \times 10^{-11}~\cite{Data}\\ 
 \end{array} \right. ,
\label{Eq:mu-decays}
\end{eqnarray}
\noindent
where ${\bar m_k}$ $\sim$ $m_k$ $\sim$ $m_{k^\prime}$ and $\xi$ estimated to be
\begin{eqnarray}
&& \xi  \sim \frac{1}{16\pi ^2 }\frac{\mu _b \mu _0 }{\bar m_k^2 }~\left( \ll 1\right)
\label{Eq:suppressed-mu-decays}
\end{eqnarray}
reads the suppression due to the approximate $L^\prime$-conservation,  
\item $\tau ^ -   ~\to~\mu^ -  e^ -  e^ + $ in Figure \ref{fig4}(a) and $\tau ^ -  ~\to~\mu^ -    \gamma$ in Figure \ref{fig4}(b) (allowed by the $L^\prime$-conservation), yielding
\begin{eqnarray}
&& \left|\frac{{ f_{\{ e\tau \} } f_{\{ e\mu\} } }}{{\bar m_k^2 }}\right| < \left\{ \begin{array}{l}
 2.1 \times 10^{ - 7}~{\rm  GeV}^{ - 2}~{\rm from}~B\left( {\tau ^ -   \to \mu^ -  e^ -  e^ +  } \right) < 1.7 \times 10^{-6} ~\cite{Data}\\ 
 4.2 \times 10^{ - 6}~{\rm  GeV}^{ - 2}~ {\rm from}~B\left( {\tau ^ -   \to \mu^ -   \gamma } \right) < 1.1 \times 10^{-6} ~\cite{Data}\\ 
 \end{array} \right. ,
\nonumber \\
&& \left|\frac{f_{[ e\tau ]} f_{[e\mu]} }{{m_h^2 }}\right| <  4.2 \times 10^{ - 6}~{\rm  GeV}^{ - 2}~ {\rm from}~B\left( {\tau ^ -   \to \mu^ -   \gamma } \right),
\label{Eq:tau-decays1}
\end{eqnarray}
\item $e^ -  e^ -   \to e^ -  e^ -  $ \cite{E_E} in Figure \ref{fig5}(a), yielding
\begin{eqnarray}
&& \left| {\frac{{f_{\{ ee\} } }}{{m_{k'} }}} \right|^2  < 4.8 \times 10^{ - 5}~{\rm  GeV}^{ - 2} ,
\label{Eq:ee-ee}
\end{eqnarray}
\item $\nu_\mu e^- \to \nu_\mu e^-$ \cite{MuDecay} in Figure \ref{fig5}(b), yielding
\begin{eqnarray}
&& \left| {\frac{{f_{[e\mu ]} }}{{m_h }}} \right|^2  < 1.7 \times 10^{ - 6}~{\rm  GeV}^{ - 2} .
\label{Eq:mu-weakdecays}
\end{eqnarray}
\end{enumerate}
It should be noted that the leading contribution of $h^+$ to $\mu^- \to e^-\gamma$, which gives 
the most stringent constraint on $h^+$, is forbidden by the $U(1)_{L^\prime}$-invariant coupling 
structure.

Typical parameter values are so chosen to satisfy these constraints:
\begin{eqnarray}
&& \left. \begin{array}{l}
 f_{[e\mu ]}  = f_{[e\tau ]}  \approx 2e \\ 
 f_{\{ ee\} }  = f_{\{ e\tau \} }  \approx e \\ 
 \end{array} \right\}{\rm ~to~suppress~higher - order~effects},
\nonumber \\
&& \left. \begin{array}{l}
 m_h  \approx 350~{\rm  GeV} \\ 
 m_k  = m_{k'}  \approx 2~{\rm  TeV} \\ 
 \mu _{\rm 0}  \approx 1.5~{\rm  TeV} \\ 
 \mu _{\rm b}  \approx {{\mu _0 } \mathord{\left/
 {\vphantom {{\mu _0 } {10}}} \right.
 \kern-\nulldelimiterspace} {10}} \\ 
 \end{array} \right\}\begin{array}{*{20}c}
   {{\rm ~to~suppress~exotic~contributions}}.
\end{array}
\label{Eq:Parameters}
\end{eqnarray}
We obtain the following numerical values:
\begin{eqnarray}
&& \left\{ \begin{array}{l}
 \Delta m_{atm}^2  \approx 2.4 \times 10^{ - 3}~{\rm  eV}^{\rm 2},  \\ 
 \Delta m_ \odot ^2  \approx 10^{ - 7}~{\rm  eV}^{\rm 2}.
 \end{array} \right.
\label{Eq:NumricalResult}
\end{eqnarray}
Therefore, we in fact successfully explain phenomena of atmospheric and solar 
neutrino oscillations characterized by $\Delta m_{atm}^2  \approx 2.4 \times 10^{ - 3} {\rm  eV}^{\rm 2}$
and $\Delta m_ \odot ^2  \approx 10^{ - 7} {\rm  eV}^{\rm 2}$.

\section{Summary}
\begin{figure}[t]
  \begin{center}
    \includegraphics*[30mm,243mm][190mm,280mm]{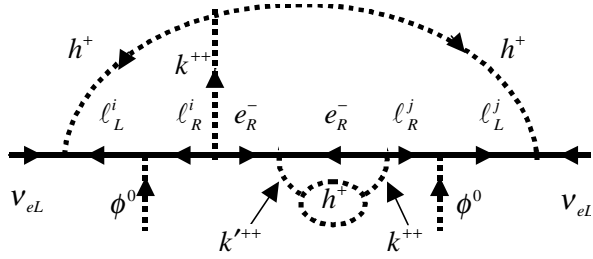}
    \caption{\label{fig6}
	Four loop-diagram for $\nu_e$-$\nu_e$.}
  \end{center}
\end{figure}

We have discussed how neutrino oscillations arise from two loop-radiative mechanism, which exhibits
\begin{enumerate}
\item bimaximal mixing due to the $L_e  - L_\mu   - L_\tau  $ conservation via the coupling of 
$e^-\tau^-k^{++}$,
\item dynamically induced tiny mass-splitting for solar neutrino oscillations due to the smallness of $m_e$ via 
$e^-e^-k^{\prime ++}$.
\end{enumerate}
The interactions required to generate two-loop Majorana neutrino masses are specified by
\begin{eqnarray}
&& \left\{ \begin{array}{l}
 f_{[ei]} \left( {\nu _{eL} \ell^i_L   - \nu ^i_L e_L^ -  } \right)h^ +   \\ 
 f_{\{ ei\} } e_R^ -  \ell^j _R  k^{ +  + }  \\ 
 \frac{1}{2}f_{\{ ee\} } e_R^ -  e_R^ -  k'^{ +  + }  \\ 
 \end{array} \right.\left( {i = \mu ,\tau } \right)  \oplus \left\{ \begin{array}{l}
 \mu _0 h^ +  h^ +  k^{ +  + \dag }  \\ 
 \mu _b h^ +  h^ +  k'^{ +  + \dag }
 \end{array} \right. .
\label{Eq:SummaryInt}
\end{eqnarray}
The resulting mass scale for neutrino masses is determined by
\begin{eqnarray}
&& \frac{{m_\tau  m_e }}{({16\pi ^2 })^2{m_k^2 }}\mu _0  \sim \frac{{m_\tau  m_e }}{({16\pi ^2 })^2{m_k }}  \sim 10^{ - 2}~{\rm  eV}.
\label{Eq:NuMassScale}
\end{eqnarray}
Thus, to obtain the neutrino mass of order of 0.01 eV is a natural consequence without fine-tuning of coupling 
parameters.  And the hierarchy of $\Delta m_ {atm} ^2 \gg \Delta m_ \odot ^2$ is expressed by the estimate 
\begin{eqnarray}
&& \Delta m_ \odot ^2  \sim \frac{{\mu _b }}{{\mu _0 }}\frac{{m_e }}{{m_\tau  }}\frac{{m_k^2 }}{{m_{k'}^2 }}\Delta m_{atm}^2,
\label{Eq:AtmSolHirearchy}
\end{eqnarray}
which ensures $\Delta m_ {atm} ^2 \gg \Delta m_ \odot ^2$  because of $m_\tau \gg m_e$.
\footnote{
One should be aware of higher-order contributions found by Lavoura in Ref.\cite{2LoopOnly}.  The (1,1)-entry of $M_\nu$, which vanishes up to the two-loop level, is induced by the four-loop diagram shown in Figure \ref{fig6}.  The contributions are at most characterized by $\delta$ $\sim$ $(16\pi^2 )^{-1}\xi m_\tau^2/m^2_k$, which should be compared with $m_e^2/m^2_{k^\prime}$.  Our parameter-setting in Eq.(\ref{Eq:Parameters}) gives $\delta$ $\sim$ 2$m_e^2/m^2_k$, which turns out to be ${\cal O}$($m_e^2/m^2_{k^\prime}$).  Therefore, our estimate of Eq.(\ref{Eq:AtmSolHirearchy}) remains valid to predict $\Delta m^2_\odot$ from $\Delta m^2_{atm}$.
}
This estimation yields the LOW solution to the solar neutrino problem.

It should be finally noted that 
\begin{itemize}
\item since the $L^\prime$-conservation forbids primary flavor-changing 
processes involving $e^-$, the coupling strengths of $h^+$ and $k^{++}$ to leptons are not severely constrained and can be as large as ${\cal O}(e)$, 
\item characteristic signatures of $h^+$ include 
\begin{eqnarray}
&& B(h^+ \to e^+{\not {E_T}}) \approx 2B(h^+ \to \mu^+{\not {E_T}} ) \approx 
2B(h^+ \to \tau^+{\not {E_T}} )
\end{eqnarray}
since $f_{[e\mu ]}$ $\approx$ $f_{[e\tau ]}$, which should be compared with \cite{AnotherHiggs}
\begin{eqnarray}
&& B(h^+ \to e^+{\not {E_T}} ) \approx B(h^+ \to \mu^+{\not {E_T}} ) \gg 
B(h^+ \to \tau^+{\not {E_T}} )
\end{eqnarray}
in the one-loop radiative mechanism with $f_{[e\mu ]}$ $\gg$ $f_{[e\tau ]}$ $\gg$ $f_{[\mu\tau ]}$ \cite{InverseHierarchy}.
\end{itemize}

\begin{center}
{\bf Acknowledgments}
\end{center}

We are grateful to M. Matsuda for his suggestion on possible constraints from $\tau$-decays.  
The work of M.Y. is supported by the Grant-in-Aid for Scientific Research No 12047223 from the Ministry of Education, Science, Sports and Culture, Japan.


\end{document}